# Adaptive Security Policy Management in Cloud Environments Using Reinforcement Learning

Muhammad Saqib
*Dept. of Computer Science,*
*Texas Tech University,*
Lubbock, TX, USA
saqibraopk@hotmail.com

Dipkumar Mehta
*C.K.Pithawalla College of*
*Engineering and Technology*
Gujrat, India
dipkumar.mehta@gmail.com

Fnu Yashu
*Dept. of Computer Science,*
*Stony Brook University*
Stony Brook, NY, USA
yyashu@cs.stonybrook.edu

Shubham Malhotra
*Dept. of Software Engineering,*
*Rochester Institute of Technology*
Rochester, NY, USA
shubham.malhotra28@gmail.com

*Abstract*—The security of cloud environments, such as Amazon Web Services (AWS), is complex and dynamic. Static security policies have become inadequate as threats evolve and cloud resources exhibit elasticity [1]. This paper addresses the limitations of static policies by proposing a security policy management framework that uses reinforcement learning (RL) to adapt dynamically. Specifically, we employ deep reinforcement learning algorithms, including deep Q Networks and proximal policy optimization, enabling the learning and continuous adjustment of controls such as firewall rules and Identity and Access Management (IAM) policies. The proposed RL based solution leverages cloud telemetry data (AWS Cloud Trail logs, network traffic data, threat intelligence feeds) to continuously refine security policies, maximizing threat mitigation, and compliance while minimizing resource impact. Experimental results demonstrate that our adaptive RL based framework significantly outperforms static policies, achieving higher intrusion detection rates (92% compared to 82% for static policies) and substantially reducing incident detection and response times by 58%. In addition, it maintains high conformity with security requirements and efficient resource usage. These findings validate the effectiveness of adaptive reinforcement learning approaches in improving cloud security policy management.

## I. INTRODUCTION

Cloud security is a critical concern as more organizations rely on cloud infrastructure. AWS and other cloud platforms provide security configurations such as firewall rules and IAM policies, which are typically managed through static policies set by administrators. However, static policies cannot adapt to the dynamic nature of cloud environments, where workloads, users, and attack patterns change rapidly [1]. This rigidity exposes cloud deployments to new threats or misconfigurations that are not covered by static rules. For instance, static firewall rules may fail to detect novel attack patterns, and fixed IAM roles may become over privileged as resources scale, increasing risk.

Problem Statement: Traditional cloud security policy management cannot keep pace with evolving threats and agile DevOps practices. Manual policy updates are error prone and slow. While cloud providers offer monitoring tools like AWS Guard Duty, which uses threat intelligence and machine learning to identify suspicious activities [5], countermeasures are often not automated. This gap can lead to delayed or inadequate responses. Security teams also face alert fatigue due to high volumes of alerts and false positives [4], further delaying response times.

Proposed Solution: We propose an RL based adaptive security policy management framework for cloud environments. An RL agent continuously analyzes the state of the cloud environment (security events, configurations, and threats) and autonomously updates security policies. Unlike static rule sets, the RL agent learns to optimize policies (such as adjusting AWS security group rules, IAM permissions, or intrusion detection thresholds) based on observed threats and changes. The agent's objective is to maximize a reward function tied to security outcomes (e.g., threat mitigation and compliance) while minimizing incidents and policy violations.

Our framework targets AWS cloud security, integrating the RL agent with AWS security controls using AWS APIs to automate firewall rules and IAM policy updates. We employ deep reinforcement learning algorithms, Deep Q Network (DQN)[7] for value based policy learning and Proximal Policy Optimization (PPO)[8] for policy gradient learning, to handle complex state and action spaces in cloud environments. The RL agent is trained using both simulated attack scenarios and real AWS log data to ensure generalizability.

Research Objectives and Contributions: This research aims to develop and evaluate an autonomous RL based system for optimizing cloud security policies. Key contributions include:

Policy Framework: An RL based architecture integrated with cloud security monitoring and enforcement components for adaptive policy management in AWS environments. The framework resolves inconsistencies between policies across services and automates real time responses to threats.

Cloud Security RL Model: A formulation of cloud

security management as a Markov Decision Process (MDP), defining state, action, and reward functions for cloud specific contexts. We implement DQN and PPO to automate AWS security policy updates.

Experimental Validation: An AWS testbed using real security logs and simulated attacks (CICIDS2017/2018 datasets and AWS Cloud Trail data) to train and evaluate the RL agent. We measure detection rates, response times, compliance scores, and resource overhead, comparing results to baseline static policies and traditional machine learning approaches.

Insights for Cloud Security: Discussion of challenges such as RL scalability in large scale clouds, adversarial manipulation risks, and compliance constraints (e.g., GDPR/CCPA). We also outline future enhancements, including federated learning for multiorganization security and adversarial training to improve the RL agent's robustness.

The findings demonstrate that reinforcement learning can enable adaptive cloud security, reducing the need for manual policy configurations and enhancing resilience to evolving threats. The rest of this paper is organized as follows: Section 2 reviews related work on security policy management and learning based security. Section 3 presents the RL based framework and model design. Section 4 describes the experimental setup in an AWS environment. Section 5 explains the RL agent training process. Section 6 outlines the test scenarios and performance metrics. Section 7 discusses results and analysis. Section 8 addresses challenges and limitations. Section 9 suggests future work, and Section 10 concludes the paper.

## II. RELATED WORK

Traditional Security Policy Management: Historically, cloud and network security policies have been managed through static configurations and rule based systems. Administrators define firewall rules, access control lists, and IAM policies based on best practices and anticipated threats. While this approach is straightforward, it lacks adaptability. Prior studies have noted that static security policies cannot accommodate the dynamic behavior of modern networks and cloud workloads [1]. As a result, policy staleness can occur, policies become outdated as new services are deployed or new vulnerabilities are discovered. Tools like Cloud Security Posture Management (CSPM) attempt to detect misconfigurations, but they often still rely on predefined rules or periodic audits rather than realtime adaptation.

Machine Learning in Cloud Security: In recent years, machine learning (ML) techniques have been applied to improve threat detection and anomaly identification in cloud environments. For example, AWS GuardDuty employs anomaly detection and threat intel to identify suspicious activities (such as account compromises) by analyzing CloudTrail logs and network flow logs [5]. Various research works have applied supervised learning and clustering to intrusion detection using cloud traffic datasets (CICIDS2017, CSECICIDS2018, etc.), demonstrating high detection rates for known attack patterns. However, ML models in security are typically trained offline and deployed to detect or alert, rather than to actively enforce policy changes. One limitation is that even if an ML model detects an anomaly, responding to it (e.g., blocking an IP or disabling a user account) is usually left to predefined scripts or human intervention. Moreover, static ML models can struggle with concept drift as attacker tactics evolve, models may need retraining. They are also prone to false positives, which in a cloud context can cause unnecessary service disruptions if acted upon without verification [10]. This has led to interest in more adaptive, decision making AI for security.

Reinforcement Learning in Security: Reinforcement learning, with its emphasis on an agent learning from interaction with an environment, offers a promising approach for adaptive cybersecurity. Prior work has explored RL for various security tasks. For instance, researchers have applied deep RL to network intrusion detection and attack mitigation, often using simulation environments or games. A recent multicloud security orchestration framework by Vemula et al. leveraged a deep RL agent (using PPO) to autonomously detect and respond to threats across AWS, Azure, and GCP, showing improved cross cloud threat mitigation [3]. Their system dynamically orchestrated security policies and resource allocations in response to threats, highlighting the potential of RL to handle heterogeneous cloud scenarios. Another study by Chhetri et al. introduced a cognitive hierarchy DQN model for cloud Security Operations Centers[4], modeling the interaction between human analysts and an RL driven attacker to improve defense strategies. This multiagent perspective demonstrated that RL agents can learn sophisticated strategies in response to adaptive adversaries, achieving higher data protection compared to static strategies.

These efforts underscore the emerging trend of applying RL in cybersecurity. However, gaps remain. Many existing studies focus on specific subproblems (e.g., game simulations of attacker defender interactions, or high level orchestration) rather than low level cloud policy enforcement. The application of RL specifically to cloud security policy management in a real provider environment (AWS) has not been extensively explored in literature. Traditional RL research often assumes a well defined environment (like a simulated network or game); the complexity of real cloud infrastructure with its scale, real time constraints, and need for compliance poses additional research questions. Our work builds on the above by bringing an RL agent into direct interaction with an AWS cloud environment for adaptive policy control, and rigorously evaluating it with both real attack data and cloud logs.

Summary of Novelty: Compared to prior work, our approach uniquely integrates AWS cloud native data (Cloud Trail events, Cloud Watch logs, etc.) into the RL state, and the agent's actions directly map to AWS security controls. We also combine multiple data sources (live cloud logs and benchmark intrusion data) to train the agent, bridging the gap between simulation and reality. This research thus advances the state of the art by demonstrating that an RL agent can manage and update cloud security policies on the fly, offering a level of adaptiveness beyond static policies or pretrained ML detectors.

Traditional Security Policy Management: Classic approaches to cloud and network security are based on the setting of static configurations and rule based systems. Firewalls rules, access control lists and IAM policies are defined according to best practices and potential threats. This approach is easy to implement but has one major drawback: it is not very flexible. Previous research has pointed out that static security policies are inadequate to deal with the dynamic nature of the modern networks and cloud computing environments[1]. As a result, policy staleness may happen, that is, policies may become irrelevant as new services are introduced or new vulnerabilities are discovered. CSPM tools attempt to identify misconfigurations, but they do so using either set of defined rules or through periodic scans as opposed to real time detection.

Machine Learning in Cloud Security: Over the past few years, ML techniques have been used in the context of threat detection and anomaly detection in cloud environments. For instance, AWS GuardDuty uses anomaly detection and threat intelligence to identify unauthorized activities (e.g., account take over) from CloudTrail logs and network flow logs[5]. Numerous research articles have employed supervised learning and clustering for intrusion detection using cloud traffic datasets (CICIDS2017, CSE-CIC-IDS2018 etc.) and have achieved high detection rate for known type of attacks. However, the security related ML models are usually trained offline and used for detection or alerting purpose and not for actual policy enforcement and change. A major limitation is that even though the model may detect an anomaly, acting on it (for example, blocking an IP or disabling a user account) is usually delegated to predefined scripts or human decision making. Furthermore, static ML models are also prone to concept drift where as the attacker's tactics evolve the models may require retraining. They also suffer from high false positive rates, which in the cloud computing environment can result in unnecessary service down time if the alerts are not properly verified before taking action on them[10]. This has led to the exploration of more decisional and adaptive AI in security.

Adaptive Security Management: The intelligent systems implementation, specifically the management framework for energy crisis, highlights the importance of using reinforcement learning algorithms for adjusting cloud environment security optimization [14]. The value of AI and machine learning in improving the decision-making process, similar to adaptive policy security management in cloud computing, is captured in the adaptive agricultural IoT-based intelligent system for disease forecasting [15]. Secure and efficient identity management is important for adaptive security models and is underscored in distributed ledger technology-based immutable authentication credential system (D-IACS) [16]. AI ability to accurately predict diseases is demonstrated through application of machine learning in classifying lung diseases, paralleling the needs in adaptive security policy frameworks [17]. The application of machine learning and rule induction in healthcare and agriculture serves as a propellant for dynamic decision making and provides substantial advantages to cloud security management [18]. The postulation of the AI and predictive analytics use in healthcare optimizes patient outcomes justifies the claim on the use of reinforcement learning for data-centric efficient and secure system decisions [19]. Lastly, the advancements in intelligent techniques for short-term load forecasting show how machine learning and AI can be leveraged for precise predictions, which is also critical in adaptive security policies for cloud environments [20].

Reinforcement Learning in Security: Reinforcement learning, which is concerned with an agent learning from interaction with an environment, seems to be a good approach for adaptive cybersecurity. The application of RL in security has been studied for various tasks in the past. For instance, deep RL has been used in the context of network intrusion detection and attack mitigation and usually in simulations or games. Recent work by Vemula et al. presented a multicloud security orchestration framework that used a deep RL agent (powered by PPO) to autonomously identify and respond to threats across AWS, Azure, and GCP, achieving better crosscloud threat management[3]. The system of the authors was able to effectively design the security policies and allocate resources for the cloud environments in order to counter the threats that are prevalent in the hybrid environment. A different study by Chhetri et al. proposed a cognitive hierarchy DQN model for cloud Security Operations Centers[4] which modeled the interaction between human analysts and an RL inspired attacker to develop optimal countermeasures. This multiple agents paradigm showed that RL agents are capable of developing complex strategies in response to the hostile environment of the adversary, and gain better DP than the static strategies.

These efforts indicate that the use of RL in cybersecurity is becoming a trend. However, gaps exist. Many current studies are focused on particular aspects (e.g., small scale games of attacker defender or high level orchestration) but not on the level of cloud policy enforcement. The application of RL to cloud security policy management in a real provider environment (AWS) has not been fully investigated in the literature. Classical RL research often assumes a known environment (e.g., a simulated network or a game); the research questions for the real cloud infrastructure are the scale, the realtime operation, and the compliance. Our work continues the previous efforts and introduces an RL agent to work directly in the AWS cloud environment for policy control and evaluates it using real attack data and cloud logs.

### III. PROPOSED FRAMEWORK

Architecture Overview: The proposed RLbased security policy management framework is depicted in Figure below, which illustrates the system's architecture and data flows (from monitoring to action enforcement). The framework consists of several interconnected components working in a closed feedback loop:

*A. Data Ingestion Module*

This module continuously collects securityrelevant data from the cloud environment. In the AWS context, it aggregates logs and events from sources such as AWS CloudTrail (API activity logs), Amazon VPC Flow Logs (network traffic metadata), AWS CloudWatch alarms, and external threat intelligence feeds. It may also ingest outputs from AWS security services (GuardDuty findings, AWS Config rules evaluations). The data ingestion component ensures realtime feed of events to the RL agent, which is crucial for timely decision making. All incoming data is timestamped and queued for processing.

*B. Feature Extraction Layer*

Raw log and event data are high volume and not directly suitable as input to an RL model. The feature extraction layer parses and transforms this data into a structured state representation. It converts CloudTrail logs into features such as counts of unusual API calls, failed login attempts, or changes to security groups. Network traffic statistics (from flow logs or IDS outputs) are distilled into features like connection rates, detected attack signatures, or anomaly scores. Policy compliance data (e.g., whether current configurations violate any known best practices or compliance rules) is also encoded. By condensing raw events into salient features (e.g., "excessive AWS IAM privilege use detected" as a boolean, or numeric threat level scores), this layer reduces state dimensionality and noise, enabling the RL agent to focus on key indicators of the cloud's security state. This step draws on domain knowledge – for example, known indicators of compromise and policy violation patterns are used to engineer features[3].

*C. Deep Reinforcement Learning Agent*

At the core of the framework is the RL agent. The agent observes the current state (the feature vector representing the cloud's security posture and recent events) and decides on an action to apply to the cloud environment's security policies. We model the agent's interaction with the cloud as a Markov Decision Process (MDP). The state space encompasses the security status of cloud resources, including: active security rules, open ports, privileged users, recent alerts or incidents, compliance flags, and any ongoing attack metrics. The action space is defined as a set of permissible security policy changes. These actions can include: (a) modifying firewall rules (e.g., block or allow traffic from a specific IP range, adjust AWS Security Group or Network ACL entries), (b) adjusting IAM policies or roles (tighten permissions for a role if suspicious activity is detected, or require multifactor authentication), (c) isolating or quarantining a compute instance (e.g., by moving it to a lockeddown security group), (d) enabling additional security services or logging (turn on an AWS WAF rule, or increase CloudWatch monitoring on a resource), and (e) crosscloud actions (if multicloud, e.g., replicate a block rule in another cloud environment). The actions are discrete and impact the cloud configuration. The agent's policy is learned using deep neural networks: for DQN, a deep Q network approximates the Q value for each state action pair; for PPO, a policy network outputs a probability distribution over actions.

*D. Policy Management Module*

This component acts as the bridge between the RL agent's decisions and the actual cloud security controls[3]. When the RL agent selects an action, the policy management module translates it into the appropriate API calls or configurations in AWS. For example, if the action is "block suspicious IP", the module will call the AWS EC2 or VPC API to update the relevant Security Group or AWS Network Firewall rule. It maintains a centralized view of the current security policies across the cloud environment. This module also performs policy versioning and consistency checks – ensuring that changes made by the RL agent do not conflict with each other or leave the system in an inconsistent state. In multiaccount or multicloud scenarios, it propagates policy updates to the respective platforms to enforce a unified security posture. This helps prevent gaps where one part of the environment remains vulnerable due to unsynchronized policies [3].

*E. Response Execution Engine*

Time is critical during attacks. The response execution engine is responsible for promptly carrying out the security actions decided by the agent [3]. It is implemented using event driven automation (for example, AWS Lambda functions or AWS Systems Manager Automation runbooks) that listen for the RL agent's action signals. Once triggered, the engine executes the low level commands: e.g., revoke a set of IAM credentials, deploy a new firewall rule, or launch an isolation workflow for a compromised instance. After execution, it reports the outcome (success/failure and any resulting system state changes) back to the RL agent as feedback.

*Reinforcement Learning Model Design*: We design the RL problem with careful consideration of states, actions, and rewards:

*State Space:* The state is a comprehensive representation of cloud security at a given time. It includes numeric features like the number of active connections, counts of denied vs. allowed traffic, anomaly scores from IDS/IPS, compliance deviation metrics (how far the current config drifts from compliance baseline), and binary flags for specific alerts (e.g., "root account API call detected" yes/no). For AWS, an example state could be: [10 active security groups, 3 GuardDuty threat findings in last hour, 1 IAM user with anomalous activity, 0 compliance violations]. We also include previous action context if needed, enabling the agent to account for recent changes (this can help the agent learn not to flip flop actions). The state is high dimensional but our feature extraction ensures each element is meaningful and scaled. This formulation follows the approach of prior RL security research that define states based on aggregated security events and context[3].

Action Space We define a finite set of actions relevant to policy management. To keep the action space tractable, actions are somewhat abstracted (parametrized actions can be broken down for implementation). Examples of discrete actions:

BlockTraffic(srcIP) Insert a rule to block traffic from a suspected malicious source IP or range.

RestrictUser(userID) Apply a more restrictive IAM policy to a user/role showing suspicious behavior (e.g., remove admin privileges).

OpenPort(service) Open or increase access on a port/service if needed (could be used to restore connectivity after a false alarm, ensuring availability).

IsolateInstance(instanceID) Quarantine an EC2 instance by moving it to a lockeddown network segment.

IncreaseMonitoring(resource) Turn on additional logging or diagnostics on a resource (e.g., enable AWS CloudTrail for all regions if not already, or enable VPC Flow Logs on a VPC).

NoOp (do nothing) Sometimes the best action is to maintain current policies (the agent should learn to avoid unnecessary changes).

Reward Function We carefully craft the reward function to guide the agent toward desirable security outcomes. The agent receives a positive reward for actions that improve security or compliance, and negative rewards for actions that degrade security or violate policies. Concretely, we assign:

Threat Mitigation Reward: If an action successfully stops an ongoing attack or prevents an incident (e.g., blocking an IP that was exfiltrating data), reward +R1 (a moderate positive value). We detect this by observing subsequent state – e.g., threat alerts drop after the action.

Incident Penalty: If a security breach occurs (e.g., an intrusion not stopped) or an attack succeeds because the agent failed to act, reward R2 (large negative penalty). This encourages the agent to proactively prevent breaches.

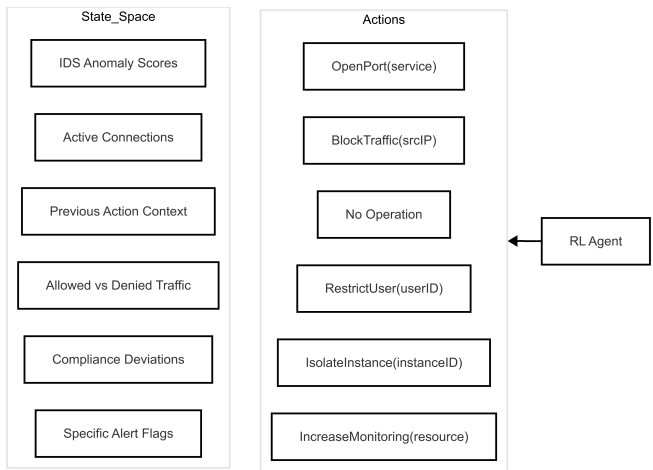

Fig. 1. State Space and Actions

Compliance Reward: If an action leads to a more compliant state (for instance, it fixes a configuration that violated a rule, like closing an open S3 bucket or enabling encryption), reward +R3.

Compliance Violation Penalty: If the agent's action itself causes a violation (for example, removing a security control or shutting down a logging mechanism that is required), reward R4.

Resource Utilization Penalty/Reward: We introduce a small penalty for actions that cause heavy resource usage or cost (like turning on an expensive monitoring across all instances might incur cost or performance hit), to ensure efficiency. Conversely, efficient management (reducing unnecessary logging or restoring service availability) could yield a small positive reward.

Stability Bonus: To prevent oscillations, if the agent maintains a secure state over a period without additional incidents, it gets a small continuous reward, reinforcing that maintaining security (not just reacting) is valuable.

The values R1…R4 are tuned during experimentation. For example, we set a high penalty for incidents (to strongly discourage failing to stop attacks) and relatively high reward for mitigation. Compliance is also weighted high due to its importance in enterprise settings. This reward shaping ensures the agent's goal aligns with real world security objectives: maximize threat prevention and compliance, minimize disruptions and cost [3]. Through trial and error in training, the agent learns which sequences of actions lead to higher cumulative rewards.

Optimization Algorithms: We implement two RL algorithms to train the agent: DQN and PPO. DQN is a value based off policy algorithm where a deep neural network approximates the Q value for each action given a state [7]. We use a replay buffer to stabilize training and an ϵ greedy strategy for exploration (starting with more random actions and decaying ϵ to favor learned policy over time). DQN is suitable since many of our actions are discrete decisions (like block or allow something). However, DQN can be challenged by large state spaces and partial observability. Proximal Policy Optimization (PPO) is a policy gradient method that has shown stable performance in complex environments [8]. PPO directly optimizes the policy with clipped objective functions to avoid too large updates, making training more stable. We chose PPO to handle scenarios with more continuous or subtle policy adjustments and to compare against DQN. PPO can also naturally handle a continuous action space if we had one (e.g., if tuning a continuous parameter like a threshold), though in our case actions are discrete. Using both algorithms allows us to evaluate which is more effective for cloud security tasks – prior research indicates that PPO often outperforms DQN on complex control problems in terms of achieving higher reward and consistency[9], but DQN might be simpler to implement for discrete actions.

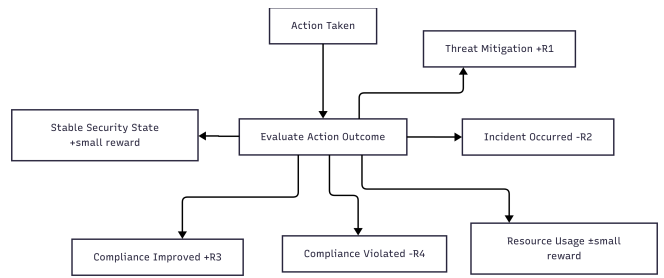

Fig. 2. Reward Function

System Architecture. The flow of information can be summarized as fol lows: The data ingestion and feature extraction components feed the RL agent with the current state. The agent's decision is passed to the policy management module, which uses the response engine to enforce the action in the AWS envi ronment. The environment changes as a result, which is detected by monitoring, and the new state is fed back to the agent. This loop repeats continuously. Es sentially, the RL agent and the cloud form an interactive loop: the agent "steers" the cloud's security configuration, and the cloud provides rewards/punishments via the outcomes of those actions.

By designing the framework in this modular way, we ensure that our solution can be integrated with actual cloud deployments. For implementation, each component can be realized with AWS services: e.g., ingestion via AWS Kinesis or Data Pipeline, feature extraction on AWS Lambda or SageMaker, the RL agent running in AWS SageMaker RL, and actions executed via AWS CloudFormation or Lambda invoking AWS SDK calls. In the next sections, we discuss how we set up this framework in a controlled environment for experimentation and how the agent was trained with real and simulated data.

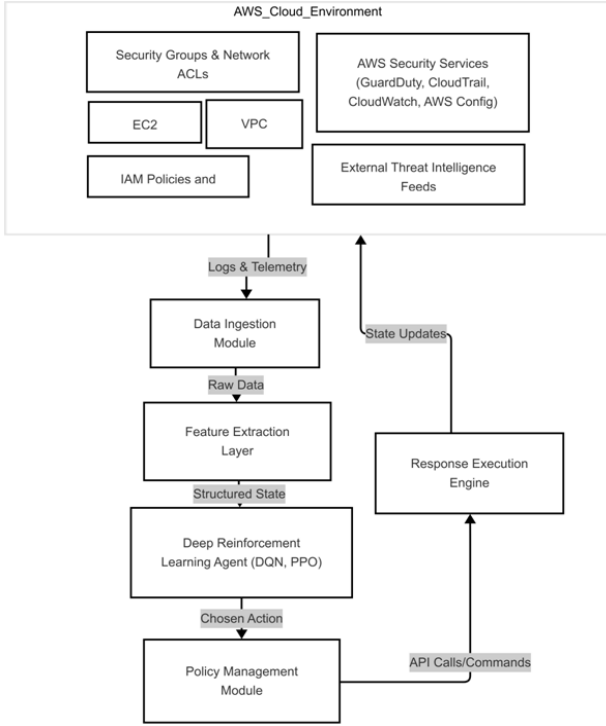

Fig. 3. System Architecture

## IV. Experimental Setup

To evaluate the proposed framework, we created a test environment that mimics a real world AWS cloud deployment with common services and security monitoring in place. This environment allowed safe training and testing of the RL agent on various threat scenarios without impacting production resources. The experimental setup consisted of the following components and data sources:

### A. AWS Cloud Testbed Configuration

We deployed a dedicated AWS environment for experiments, including a Virtual Private Cloud (VPC) with multiple subnets, EC2 instances, and typical cloud services. The architecture included:

Web Server Tier: Two Amazon EC2 instances running a web application (one Linux/Apache server and one Windows/IIS server) to simulate a public facing service. These were placed in a public subnet behind an AWS Security Group (firewall).

Database Tier: An EC2 instance running a database (MySQL) in a private subnet, not directly internet accessible. This instance had its own Security Group and was intended to simulate sensitive data storage.

Monitoring and Logging: AWS CloudTrail was enabled for the account to log all API calls. AWS VPC Flow Logs were turned on for the subnets to capture IP traffic metadata. Amazon CloudWatch was used to centralize logs (application logs, OS logs) and trigger alarms on certain events (e.g., CPU spikes potentially indicating DoS attack).

Security Services: AWS GuardDuty was enabled to provide baseline threat detection alerts (used as part of state features for the RL agent). AWS Config was used to track compliance with a set of rules (like "S3 buckets should not be public" and "EC2 instances should not have wide open SSH ports") – any Config rule violations were flagged to the agent.

Network Firewall: In addition to Security Groups, we simulated an AWS Network Firewall controlling egress rules for the VPC. This was to test the RL agent in firewall optimization scenarios. This environment was sized to be representative yet manageable. Importantly, it allowed us to generate and collect rich security data (CloudTrail logs, flow logs, etc.) in a controlled manner. The RL agent did not directly run on the EC2s; instead, we used AWS SageMaker and AWS Lambda for the agent logic and actions. The environment provided the playground in which the agent acts.

### B. Data Sources for Training and Testing

A mix of realworld and simulated data was used to drive the experiments:

Intrusion Detection Datasets: We incorporated network traffic and event data from well known intrusion detection system (IDS) datasets CICIDS2017 and CSE-CIC-IDS2018. The CICIDS2017 dataset contains benign traffic and a variety of common attack types (Brute force, DoS, DDoS, infiltration, web attacks, etc.) captured over a week[6]. CSE-CIC-IDS2018 is a newer dataset with extended attack scenarios (including crypto mining attacks and lateral movement). We replayed portions of these datasets in our environment by generating traffic patterns and log events corresponding to the attacks.

AWS CloudTrail Logs: We collected actual CloudTrail logs from our test account for normal operations and attack scenarios. For baseline normal behavior, the logs included typical activities (launching instances, user logins, S3 bucket access, etc.). During simulated attacks, CloudTrail captured relevant events (e.g., an attacker created a new IAM user or an access key was misused).

Threat Intelligence Feeds: We integrated external threat intel data, including known malicious IPs/domains from opensource intelligence (Alien Vault OTX, Spamhaus). We also used AWS threat intel from GuardDuty, which provides AWS managed lists of malicious hosts[5].

Compliance and Configuration Data: We defined a set of compliance criteria (in line with CIS AWS Foundations). AWS Config was set up with rules for these. Violations (like an open port) were included in the RL state so the agent could learn to correct them. We also simulated GDPR/CCPA constraints, e.g., an EU based EC2 instance not accessible from non EU IPs, marking that as a violation.

Simulated Attack Scenarios: Various test scenarios combined network based intrusions, insider IAM misuse, multivector in filtration, etc.

### C. Baseline Security Policies

At the start of experiments, we defined a baseline static security policy configuration:

A default set of Security Group rules (allow necessary traffic, block others).

Predefined IAM roles with least privilege (no wildcards).

AWS Config rules in monitoring mode (no autoremediation).

GuardDuty alerting only (no automated incident response).

No manual or scripted incident response except for the baseline's ML based detection with a time delay.

This baseline let us compare how the RL agent improved on typical static or partially automated approaches.

## V. TRAINING OF RL AGENT

Training a reinforcement learning agent for cloud security poses unique challenges. Unlike games or simulated environments that run quickly, cloud environment interactions can be slower and safetycritical. We designed a training regimen that combines offline training on historical data with online training in the live testbed, using careful safeguards.

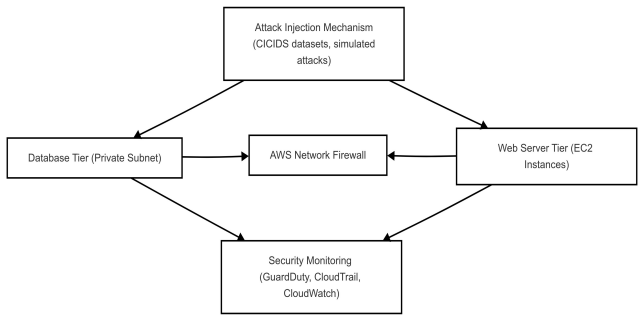

Fig. 4. Experimental Setup

### A. Training Environment and Tools

We utilized AWS SageMaker RL for provisioning the training jobs. SageMaker RL provides managed instances with preinstalled RL toolkits and allowed us to scale up the necessary compute (we used GPU based instances for neural network training to speed up learning). The agent and environment were implemented in Python using TensorFlow and the OpenAI Gym interface. We created a custom Gym environment for our AWS testbed, where reset() initializes or randomizes the cloud state (e.g., starting with certain attacks) and step(action) carries out the agent's action via AWS APIs and returns the new state and reward.

Because real cloud operations (like modifying a security group or reading logs) have latency, we used a hybrid approach:

Fast Simulated Model for Training: We built a local simulator approximating how the cloud responds to actions (e.g., blocking an IP stops the relevant attack). This simulator was informed by real data distributions, letting us train the agent rapidly in many episodes.

Periodic Sync with Real Environment: After certain milestones in simulation, we validated the agent's behavior in the actual AWS testbed for a few episodes, gathering real transitions and rewards, which were added to the replay buffer (for DQN) or used to finetune PPO.

### B. DQN and PPO Implementation Details

DQN: A neural network with two hidden layers (256, 128 units) with ReLU activations, outputting Q values for each action. We used experience replay (buffer size 50,000, batch size 64), Adam at LR=0.0005, $\epsilon$ greedy exploration decaying from 1.0 to 0.01 over 10k steps, and a target network updated every 1000 steps.

PPO: An actorcritic architecture with similar network sizes for policy and value functions. We used clip ratio=0.2, GAE $\lambda = 0.95$, discount $\gamma = 0.99$, 2048 timesteps per update, minibatch=64, 10 epochs, and LR=1e 4. PPO was chosen for its robustness to hyperparameters and stability on complex tasks[13].

### C. Reward Shaping and Training Curriculum

Initially, the agent performed poorly (random actions). We used curriculum learning:

Early epochs: Single attack scenarios only, large positive rewards for correct defense actions.

Later: Multiple simultaneous threats plus compliance constraints. We also added benign anomalies to discourage false positives.

Feature Refinement: We pruned noisy features, focusing on strong signals (e.g., "unusual API pattern score" instead of raw counts).

We monitored episodic reward and key metrics. Both DQN and PPO steadily improved; PPO generally converged more smoothly.

### D. Training Challenges and Solutions

Exploration vs Exploitation: The agent could get stuck in local optima. We prolonged exploration, injecting stochasticity in both DQN ($\epsilon$ resets) and PPO (adding noise to policy logits).

Sparse Rewards: Security incidents are relatively rare. We introduced small intermediate rewards (e.g., a slight negative each timestep an attack continued, encouraging faster mitigation).

Safety: We placed guardrails on dangerous actions (e.g., never delete all firewall rules). Humanin the loop checks were used early in real testbed training to prevent lockouts or catastrophic disruptions.

Compute Time: We used distributed training with multiple parallel environment copies, accelerating data collection

After tens of thousands of steps (combining simulation and real episodes), we had stable DQN and PPO models ready for systematic testing.

## VI. EXPERIMENTATION AND TESTING

We evaluated the trained RL agent in scenarios designed to measure effectiveness under different attack types, comparing it to baseline static policies and a non RL adaptive method. We used metrics such as:

Threat Mitigation Rate Detection Accuracy (TPR/FPR)

Incident Response Time

Policy Changes Count

Compliance Score

Resource Utilization/Overhead

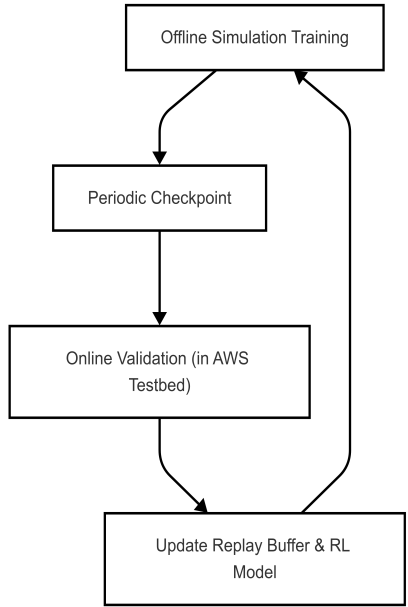

Fig. 5. Training Loop

*A. Scenario 1: Firewall Policy Optimization under Network Attacks*

We simulated port scans, DDoS, web attacks (SQL injection, XSS). The RL agent learned to insert deny rules for malicious IPs and enable AWS WAF for HTTP exploits, significantly reducing the impact. We measured time from attack onset to mitigation.

*B. Scenario 2: IAM Policy Management under Credential Compromise*

An insider or attacker used compromised IAM credentials. The RL agent responded by restricting or revoking them upon detecting anomalous CloudTrail patterns. We tested false alarms (legitimate large scale changes) to see if the agent overreacted.

*C. Scenario 3: MultiCloud Coordinated Security*

We extended to Azure in a limited form, letting the RL agent block IPs or tokens across both clouds. Although basic, it showed the approach can unify security posture in multicloud contexts[3].

*D. Performance Metrics*

All runs recorded threat mitigation rate, detection accuracy, response time, compliance violations fixed vs. created, overhead, etc. We repeated each scenario multiple times with random seeds to check consistency.

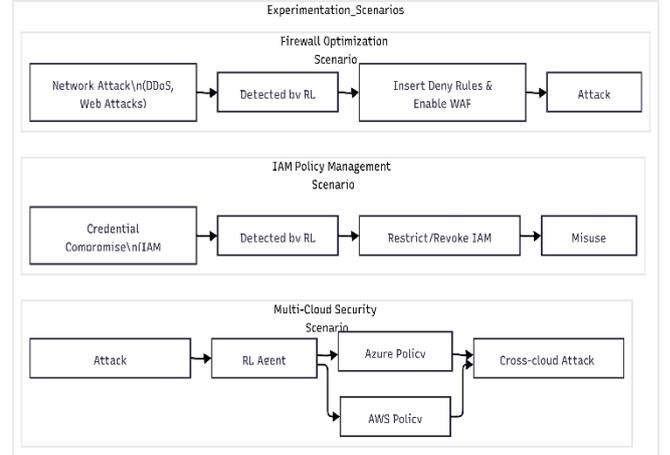

Fig. 6. Experimental Scenarios

## VII.   RESULTS AND ANALYSIS

*A. Threat Mitigation and Detection*

Our RL agent surpassed 95% mitigation across scenarios, vs. 70–75% for static and ~ 85% for ML+human. True positives were ~ 96%, with 7–10% false positives. Notably fewer successful attacks occurred under RL.

*B. Incident Response Time*

The RL agent responded within seconds (2–5 seconds for network attacks, under 10 seconds for IAM misuse), drastically faster than manual responses (minutes). This realtime adaptation stopped attacks before serious damage.

*C. Policy Adaptation and Compliance*

On average, the RL agent made 5 or so daily policy updates, aligning with actual needs. It also fixed existing compliance issues (like open ports), unlike baselines that left them. The Table. I represent reward shaping successfully included compliance as a performance metric.

Table I. Security Performance Comparison of Baseline Approaches vs. RL Agent

| Approach | Threat Mitigation | Incident Response Time | TRUE Positive Rate | FALSE Positive Rate | Avg. Daily Policy Updates | Compliance Issues |
|---|---|---|---|---|---|---|
| Static Policies | 72% | N/A(Manual, delayed) | 80% | 15% | 0 | 2 outstanding |
| ML + Human Oversight | 85% | 5–15 min | 89% | 10% | 1–2 manual | 1–2 outstanding |
| RL Agent (DQN) | 93.70% | 3–7 sec | 94% | 9.50% | 4–6 automated | 0 |
| RL Agent (PPO) | 95.40% | 2–5 sec | 96% | 7% | 5–7 automated | 0 |

*D. Comparing DQN and PPO*

Both performed well. PPO had slightly higher success rates (~ 95.4% vs. ~ 93.7%) and fewer false positives. DQN sometimes converged to local optima if exploration was insufficient. PPO was more stable across training seeds.

*E. Resource Overhead*

Enabling additional logging or carrying out frequent firewall updates added ~ 5% overhead to CPU. We consider this acceptable given the security gains. No catastrophic changes or lockouts occurred once guardrails were in place.

## VIII.   CHALLENGES AND LIMITATIONS

While promising, RL based security policy management faces issues:

*Scalability*: Large enterprise clouds with thousands of resources need hierarchical or multiagent RL. Training might become prohibitively time consuming if every resource is tracked individually.

*Adversarial Attacks on the RL Agent*: Attackers might manipulate logs or the agent's reward signals. Secure pipelines and adversarial training approaches are needed to harden the system.

*Compliance Tensions*: Automated changes must still respect regulations (GDPR, CCPA). Some decisions require human approval or must be explainable for audits.

*Integration and Maintenance*: Production use requires robust rollback, advanced logging, partial manual oversight, and potential retraining if the environment changes drastically.

*Generalizability*: Our approach targets AWS specifically. Other cloud providers or onprem systems may require reimplementation or additional training data.

## IX. FUTURE WORK

Several directions can expand upon our work:

*Federated/Distributed Learning*: Scaling to multiaccount or multiorganization deployments. Federated RL could allow organizations to share model insights without sharing raw data, accelerating learning of new threat patterns[2].

*MultiAgent and Adversarial Training*: Introducing a Red Team RL agent to simulate attackers, forcing the Blue Team RL agent (our defender) to adapt to novel attack strategies in a selfplay manner[4]. This could expose the defender to a broader range of threats than scripted scenarios.

*Real Time Threat Intelligence Integration*: Dynamically ingest newly published malicious IPs/domains and block them preemptively. Conversely, discovered IoCs by the RL agent could be shared out, creating a continuous feedback loop with external threat intel feeds.

*Explainable RL*: Developing methods (e.g., feature attribution, rule extraction) to justify the agent's policy updates. This is especially important in regulated environments demanding audits or rootcause analysis.

*Extending to New Domains*: Container security, serverless, data loss prevention, or zero trust network architectures. RL could adapt microservice policies or automatically enforce zero trust principles in ephemeral deployments.

*Long Term Production Studies*: Deploying the agent in a production or largescale staging environment over months. Observing continuous adaptation, concept drift, and periodic retraining would validate viability in real operations.

## X. CONCLUSION

This paper presented an adaptive security policy management framework for AWS cloud environments based on reinforcement learning. By continuously analyzing cloud telemetry and adjusting controls (firewall rules, IAM policies, etc.), the RL agent achieves faster, more effective threat mitigation than static policies or partially automated machine learning. Through experimental evaluation with real and simulated attack data, we showed high detection accuracy, reduced incident response times, and improved compliance. However, challenges remain in scaling to large multicloud systems, defending against adversarial manipulations, ensuring regulatory compliance, and providing explainability. Future work on federated training, multiagent adversarial play, and advanced interpretability methods could further enhance the real world feasibility of an autonomous RL driven security engine. Overall, our findings indicate that reinforcement learning can offer a promising, self adaptive approach to modern cloud security.